%% file: Template.tex
\documentclass{article}
\usepackage[preprint]{spconf}
\usepackage{amsmath,graphicx,amssymb,amsfonts}
\usepackage{xcolor}
\usepackage{pgfplots}
\pgfplotsset{compat=1.17} 
\usepackage{tikz}  
\usepackage{tikz-3dplot} 
\usetikzlibrary{shapes,arrows, positioning, automata, shadows,arrows.meta,backgrounds,fit, calc}
\usepackage{tkz-fct, tikz-dimline}
\usepackage{multirow}
\usepackage{subcaption}
\usepackage[hidelinks]{hyperref}

\tikzset{cross/.style={cross out, draw=black, minimum size=2*(#1-\pgflinewidth), inner sep=0pt, outer sep=0pt},cross/.default={4pt}}

\copyrightnotice{\fbox{\parbox{\dimexpr\textwidth-\fboxsep-\fboxrule\relax}{\footnotesize \textcopyright 2022 IEEE. Personal use of this material is permitted.
			Permission from IEEE must be obtained for all other uses, in any current or future
			media, including reprinting/republishing this material for advertising or promotional
			purposes, creating new collective works, for resale or redistribution to servers or
			lists, or reuse of any copyrighted component of this work in other works.
			DOI: \href{https://doi.org/10.1109/ICIP46576.2022.9897920}{10.1109/ICIP46576.2022.9897920}
}}}

\title{Frequency-Selective Geometry Upsampling of Point Clouds}
\name{Viktoria Heimann, Andreas Spruck, and André Kaup }
\address{Multimedia Communications and Signal Processing\\
Friedrich-Alexander-Universit\"at Erlangen-N\"urnberg}

\begin{document}

\ninept
\maketitle
%
\begin{abstract}
The demand for high-resolution point clouds has increased throughout the last years. However, capturing high-resolution point clouds is expensive and thus, frequently replaced by upsampling of low-\mbox{resolution} data. Most state-of-the-art methods are either restricted to a rastered grid, incorporate normal vectors, or are trained for a single use case. We propose to use the frequency selectivity \mbox{principle}, where a frequency model is estimated locally that approximates the surface of the point cloud. Then, additional points are inserted into the approximated surface. Our novel frequency-selective \mbox{geometry} upsampling shows superior results in terms of subjective as well as objective quality compared to state-of-the-art methods for scaling factors of \(2\) and \(4\). On average, our proposed method shows a \(4.4\)~times smaller point-to-point error than the second best state-of-the-art PU-Net for a scale factor of \(4\).

\end{abstract}
\section{Introduction and Related Work}
\label{sec:intro}
\input{introduction}
\section{Proposed Framework}
\label{sec:proposed}
\input{framework}

\section{Evaluation}
\label{sec:eval}
\input{evaluation}

\section{Conclusion}
\label{sec:conclusion}
In this paper, we proposed Frequency-Selective Geometry Upsampling (FSGU) for point cloud upsampling. The proposed technique approximates the object's surface locally using DCT basis functions. It is not dependent on additional information such as normal vectors nor restricted to a rastered grid. For the local surface approximation, a frequency model is estimated iteratively yielding a continuous representation of the surface. FSGU clearly outperforms current state-of-the-art methods in both, metric-based objective and visual subjective, quality. FSGU yields a 4.4 times smaller point to point error than the second best method, PU-Net.

\section{Acknowledgment}
This work was partly funded by the Deutsche Forschungsgemeinschaft (DFG, German Research Foundation) – SFB 1483 – Project-ID 442419336, EmpkinS.

%



\newpage
\bibliographystyle{IEEEbib}
\label{sec:ref}
\bibliography{bib_4icip2022}

\end{document}

%% file: introduction.tex

The growing demand of 3D captures of our environment increases the importance of point cloud data. Point clouds are composed of single points in three-dimensional space. Each point can be located at any arbitrary position. Furthermore, each point may hold an associated attribute such as color, normal vector, or texture.\par
Point clouds are often used for tele-immersive videos or in virtual and augmented reality applications \cite{Mekuria_2017, Held_2012}. Furthermore, autonomous vehicles use Light Detection And Ranging (LiDAR) sensors to model the environment of the vehicle \cite{Chen_2017}. LiDAR is also used in construction engineering and archaeology to obtain a plot of a building or heritage sites and objects \cite{Mahmood_2020, Andriasyan_2020}. All these applications require high-quality capturing of the point clouds. However, the acquisition of point clouds with high resolution is an expensive task. Thus, increasing the resolution of a point cloud using a post processing step is a suitable alternative. \par
As the point positions are not restricted to a rastered grid, the first task for point cloud upsampling is to define the positions of the newly generated points and add additional points into the point cloud. We refer to the challenge of inserting additional points into the surface of the point cloud as \textit{geometry upsampling} in the following. A related problem to geometry upsampling is point cloud reconstruction, where additional points are inserted to fill holes and reconstruct missing parts in the object's surface. First approaches are based on solving an indicator function in three-dimensional space by e.g. applying a Fast Fourier Transform (FFT) \cite{Kazhdan_2005}. This solution has the drawback that FFT requires a regular grid with a resolution of a power of two and, thus, it is not widely applicable. Various approaches were developed that are using point set surfaces. In order to guarantee smooth surfaces, local areas of the surfaces have to be up- and downsampled. Alexa et al. \cite{Alexa_2001} were the first who applied a Voronoi tessellation and inserted additional points in the point cloud at the center positions of the Voronoi cells.\par
A widely used approach for point cloud upsampling is Edge-Aware Resampling (EAR) \cite{Huang_2013}. It incorporates normal vectors in the upsampling scheme. A point cloud is first resampled away from the edges, as it is assumed that captured data and estimated normals are more accurate away from edges. Then, the remaining regions are upsampled. Due to the unordered saving of points in a point cloud data type, links between neighboring points are not established implicitly. This is a great challenge for the processing of point clouds. Thus, Dinesh et al. \cite{Dinesh_2019} tried to overcome this issue by creating a k-nearest-neighbor graph first. For the upsampling of a point cloud, they incorporate normal vectors, too. Furthermore, they assume the point cloud's surface to be piecewise smooth. Hence, additional points are inserted at the center position of the triangles generated by a Delaunay triangulation. Then, the final position is refined by formulating the problem as a minimization of a graph-total variation. However, normal vectors are not available for all point clouds and the computation of normal vectors might yield inaccuracies especially at edges. Furthermore, the computation of normal vectors is highly sensitive to noise which often occurs in point clouds in the capturing process.\par
In recent years, also data-driven approaches for point cloud upsampling were developed. A pioneer in this field was PU-Net \cite{Yu_2018_PUNet}. It uses extracted patches of the point cloud, extracts multi-level features and subsequently expands these to a higher resolution. In addition, a joint loss function is used that balances between uniform distribution of the points and a smooth surface. Ongoing work uses the structure of PU-Net hierarchically \cite{Yifan_2019} or combines it with a discriminator and generator structure in PU-GAN \cite{Li_2019_PUGan}. EC-Net focuses on upsampling the edge regions of point clouds \cite{Yu_2018_ECNet}. PUGeoNet \cite{Li_2021} uses a different approach as it parametrizes the three-dimensional surface and transforms it to a two-dimensional domain. The surface is expanded and the additional points are inserted. Finally, the points are retransformed to the three-dimensional space.\par  
The disadvantage of these data-driven approaches is that they only show high qualities for the data set and scale factor that they are trained for. A high-quality method that upsamples a low-resolution point cloud by any arbitrary scale factor without the aid of additional information such as normal vectors or a fixed rastered grid has not yet been reported. Thus, the objective of this work was to develop a technique that is able to upsample a point cloud by any arbitrary scale factor and is not dependent on normal vectors or a voxelization of the point cloud.\par
Our proposed frequency-selective geometry upsampling framework is presented in the upcoming section. Thereafter in Section \ref{sec:eval} follows an extensive evaluation of our framework. Finally, the paper closes with a conclusion in Section \ref{sec:conclusion}.

\begin{figure*}[t!]
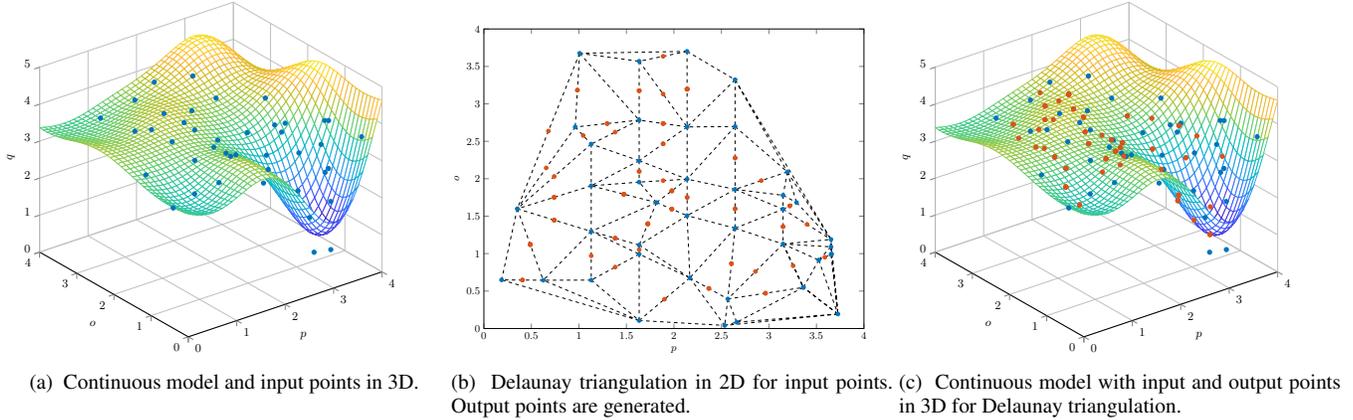

\begin{subfigure}[t]{.33\textwidth}
\scalebox{.5}{\input{figures/Duck_ModelWithInput}}
\caption{\label{Fig:duckmodelinput} Continuous model and input points in 3D.}
\end{subfigure}
\begin{subfigure}[t]{.33\textwidth}
\scalebox{.44}{\input{figures/Duck_GenerateOutput}}
\caption{\label{Fig:inputoutput} Delaunay triangulation in 2D for input points. Output points are generated.}
\end{subfigure}
\begin{subfigure}[t]{.33\textwidth}
\scalebox{.5}{\input{figures/Duck_ModelWithInAndOutput}}
\caption{\label{Fig:duckmodelinputoutput} Continuous model with input and output points in~3D for Delaunay triangulation.}
\end{subfigure}
\caption{\label{Fig:modelling}Model generation process for a block of the \texttt{Duck} point cloud. Blue points are the original points, red points are the upsampled points. The continuous model is depicted as a mesh plot.}
\end{figure*}   

%% file: figures/Duck_GenerateOutput.tex
%
\definecolor{mycolor1}{rgb}{0.00000,0.44700,0.74100}%
\definecolor{mycolor2}{rgb}{0.85000,0.32500,0.09800}%
\begin{tikzpicture}

\begin{axis}[%
width=4.521in,
height=3.566in,
at={(0.758in,0.481in)},
scale only axis,
unbounded coords=jump,
xmin=0,
xmax=4,
xlabel={\(p\)},
ymin=0,
ymax=4,
ylabel={\(o\)},
axis background/.style={fill=white},
legend style={legend cell align=left, align=left, draw=white!15!black}
]
\addplot [color=black, dashed]
  table[row sep=crcr]{%
2.13809967041016	3.70050048828125\\
2.64369964599609	3.32180404663086\\
nan	nan\\
2.13809967041016	3.70050048828125\\
1.00749969482422	3.67480087280273\\
nan	nan\\
2.13809967041016	3.70050048828125\\
1.63420104980469	3.56869888305664\\
nan	nan\\
2.13809967041016	3.70050048828125\\
2.13899993896484	2.69209671020508\\
nan	nan\\
2.64369964599609	3.32180404663086\\
2.64369964599609	2.69680023193359\\
nan	nan\\
2.64369964599609	3.32180404663086\\
2.13899993896484	2.69209671020508\\
nan	nan\\
2.64369964599609	3.32180404663086\\
3.19409942626953	2.09109878540039\\
nan	nan\\
2.64369964599609	3.32180404663086\\
3.65320014953613	1.189697265625\\
nan	nan\\
1.00749969482422	3.67480087280273\\
1.63420104980469	3.56869888305664\\
nan	nan\\
1.00749969482422	3.67480087280273\\
1.63420104980469	2.78250122070312\\
nan	nan\\
1.00749969482422	3.67480087280273\\
0.961099624633789	2.69100189208984\\
nan	nan\\
1.00749969482422	3.67480087280273\\
0.35099983215332	1.59850311279297\\
nan	nan\\
1.63420104980469	3.56869888305664\\
2.13899993896484	2.69209671020508\\
nan	nan\\
1.63420104980469	3.56869888305664\\
1.63420104980469	2.78250122070312\\
nan	nan\\
2.64369964599609	2.69680023193359\\
2.13899993896484	2.69209671020508\\
nan	nan\\
2.64369964599609	2.69680023193359\\
3.19409942626953	2.09109878540039\\
nan	nan\\
2.64369964599609	2.69680023193359\\
2.64369964599609	1.85800170898438\\
nan	nan\\
2.64369964599609	2.69680023193359\\
2.13899993896484	1.99480056762695\\
nan	nan\\
2.13899993896484	2.69209671020508\\
1.63420104980469	2.78250122070312\\
nan	nan\\
2.13899993896484	2.69209671020508\\
1.63420104980469	2.24029922485352\\
nan	nan\\
2.13899993896484	2.69209671020508\\
2.13899993896484	1.99480056762695\\
nan	nan\\
1.63420104980469	2.78250122070312\\
0.961099624633789	2.69100189208984\\
nan	nan\\
1.63420104980469	2.78250122070312\\
1.63420104980469	2.24029922485352\\
nan	nan\\
1.63420104980469	2.78250122070312\\
1.1294994354248	2.46269607543945\\
nan	nan\\
0.961099624633789	2.69100189208984\\
1.1294994354248	2.46269607543945\\
nan	nan\\
0.961099624633789	2.69100189208984\\
0.35099983215332	1.59850311279297\\
nan	nan\\
1.63420104980469	2.24029922485352\\
2.13899993896484	1.99480056762695\\
nan	nan\\
1.63420104980469	2.24029922485352\\
1.1294994354248	2.46269607543945\\
nan	nan\\
1.63420104980469	2.24029922485352\\
1.63420104980469	1.95490264892578\\
nan	nan\\
1.63420104980469	2.24029922485352\\
1.1294994354248	1.90569686889648\\
nan	nan\\
3.19409942626953	2.09109878540039\\
2.64369964599609	1.85800170898438\\
nan	nan\\
3.19409942626953	2.09109878540039\\
3.14849853515625	1.77890014648438\\
nan	nan\\
3.19409942626953	2.09109878540039\\
3.2903995513916	1.68230056762695\\
nan	nan\\
3.19409942626953	2.09109878540039\\
3.65320014953613	1.189697265625\\
nan	nan\\
2.64369964599609	1.85800170898438\\
2.13899993896484	1.99480056762695\\
nan	nan\\
2.64369964599609	1.85800170898438\\
3.14849853515625	1.77890014648438\\
nan	nan\\
2.64369964599609	1.85800170898438\\
2.64369964599609	1.33929824829102\\
nan	nan\\
2.64369964599609	1.85800170898438\\
2.13899993896484	1.50730133056641\\
nan	nan\\
2.64369964599609	1.85800170898438\\
3.14849853515625	1.59299850463867\\
nan	nan\\
2.13899993896484	1.99480056762695\\
1.81710052490234	1.68230056762695\\
nan	nan\\
2.13899993896484	1.99480056762695\\
1.63420104980469	1.95490264892578\\
nan	nan\\
2.13899993896484	1.99480056762695\\
2.13899993896484	1.50730133056641\\
nan	nan\\
1.1294994354248	2.46269607543945\\
1.1294994354248	1.90569686889648\\
nan	nan\\
1.1294994354248	2.46269607543945\\
0.35099983215332	1.59850311279297\\
nan	nan\\
1.81710052490234	1.68230056762695\\
1.63420104980469	1.95490264892578\\
nan	nan\\
1.81710052490234	1.68230056762695\\
2.13899993896484	1.50730133056641\\
nan	nan\\
1.81710052490234	1.68230056762695\\
1.1294994354248	1.90569686889648\\
nan	nan\\
1.81710052490234	1.68230056762695\\
1.63420104980469	1.11449813842773\\
nan	nan\\
1.81710052490234	1.68230056762695\\
1.1294994354248	1.29710006713867\\
nan	nan\\
1.63420104980469	1.95490264892578\\
1.1294994354248	1.90569686889648\\
nan	nan\\
3.14849853515625	1.77890014648438\\
3.14849853515625	1.59299850463867\\
nan	nan\\
3.14849853515625	1.77890014648438\\
3.2903995513916	1.68230056762695\\
nan	nan\\
2.64369964599609	1.33929824829102\\
2.13899993896484	1.50730133056641\\
nan	nan\\
2.64369964599609	1.33929824829102\\
3.14849853515625	1.59299850463867\\
nan	nan\\
2.64369964599609	1.33929824829102\\
2.16760063171387	0.676101684570312\\
nan	nan\\
2.64369964599609	1.33929824829102\\
3.14849853515625	1.12820434570312\\
nan	nan\\
2.64369964599609	1.33929824829102\\
2.56709861755371	0.394302368164062\\
nan	nan\\
2.13899993896484	1.50730133056641\\
1.63420104980469	1.11449813842773\\
nan	nan\\
2.13899993896484	1.50730133056641\\
2.16760063171387	0.676101684570312\\
nan	nan\\
1.1294994354248	1.90569686889648\\
0.35099983215332	1.59850311279297\\
nan	nan\\
1.1294994354248	1.90569686889648\\
1.1294994354248	1.29710006713867\\
nan	nan\\
1.63420104980469	1.11449813842773\\
2.16760063171387	0.676101684570312\\
nan	nan\\
1.63420104980469	1.11449813842773\\
1.1294994354248	1.29710006713867\\
nan	nan\\
1.63420104980469	1.11449813842773\\
1.63420104980469	0.993598937988281\\
nan	nan\\
3.14849853515625	1.59299850463867\\
3.2903995513916	1.68230056762695\\
nan	nan\\
3.14849853515625	1.59299850463867\\
3.14849853515625	1.12820434570312\\
nan	nan\\
3.14849853515625	1.59299850463867\\
3.65320014953613	1.189697265625\\
nan	nan\\
3.2903995513916	1.68230056762695\\
3.65320014953613	1.189697265625\\
nan	nan\\
2.16760063171387	0.676101684570312\\
1.63420104980469	0.993598937988281\\
nan	nan\\
2.16760063171387	0.676101684570312\\
2.56709861755371	0.394302368164062\\
nan	nan\\
2.16760063171387	0.676101684570312\\
1.63420104980469	0.106800079345703\\
nan	nan\\
2.16760063171387	0.676101684570312\\
2.53399848937988	0.0407981872558594\\
nan	nan\\
0.35099983215332	1.59850311279297\\
1.1294994354248	1.29710006713867\\
nan	nan\\
0.35099983215332	1.59850311279297\\
0.624700546264648	0.644702911376953\\
nan	nan\\
0.35099983215332	1.59850311279297\\
0.186100006103516	0.649101257324219\\
nan	nan\\
1.1294994354248	1.29710006713867\\
1.63420104980469	0.993598937988281\\
nan	nan\\
1.1294994354248	1.29710006713867\\
1.1294994354248	0.64630126953125\\
nan	nan\\
1.1294994354248	1.29710006713867\\
0.624700546264648	0.644702911376953\\
nan	nan\\
1.63420104980469	0.993598937988281\\
1.1294994354248	0.64630126953125\\
nan	nan\\
1.63420104980469	0.993598937988281\\
1.63420104980469	0.106800079345703\\
nan	nan\\
3.14849853515625	1.12820434570312\\
3.65320014953613	1.09080123901367\\
nan	nan\\
3.14849853515625	1.12820434570312\\
2.56709861755371	0.394302368164062\\
nan	nan\\
3.14849853515625	1.12820434570312\\
3.65320014953613	1.189697265625\\
nan	nan\\
3.14849853515625	1.12820434570312\\
3.52409934997559	0.914199829101562\\
nan	nan\\
3.14849853515625	1.12820434570312\\
3.36020088195801	0.547897338867188\\
nan	nan\\
3.14849853515625	1.12820434570312\\
3.36400032043457	0.550399780273438\\
nan	nan\\
3.65320014953613	1.09080123901367\\
3.6599006652832	0.994895935058594\\
nan	nan\\
3.65320014953613	1.09080123901367\\
3.65320014953613	1.189697265625\\
nan	nan\\
3.65320014953613	1.09080123901367\\
3.52409934997559	0.914199829101562\\
nan	nan\\
3.65320014953613	1.09080123901367\\
3.65320014953613	0.989299774169922\\
nan	nan\\
3.6599006652832	0.994895935058594\\
3.65320014953613	1.189697265625\\
nan	nan\\
3.6599006652832	0.994895935058594\\
3.72599983215332	0.19329833984375\\
nan	nan\\
3.6599006652832	0.994895935058594\\
3.65320014953613	0.989299774169922\\
nan	nan\\
2.56709861755371	0.394302368164062\\
3.36020088195801	0.547897338867188\\
nan	nan\\
2.56709861755371	0.394302368164062\\
2.53399848937988	0.0407981872558594\\
nan	nan\\
2.56709861755371	0.394302368164062\\
2.6609001159668	0.084503173828125\\
nan	nan\\
1.1294994354248	0.64630126953125\\
1.63420104980469	0.106800079345703\\
nan	nan\\
1.1294994354248	0.64630126953125\\
0.624700546264648	0.644702911376953\\
nan	nan\\
1.63420104980469	0.106800079345703\\
0.624700546264648	0.644702911376953\\
nan	nan\\
1.63420104980469	0.106800079345703\\
2.53399848937988	0.0407981872558594\\
nan	nan\\
1.63420104980469	0.106800079345703\\
0.186100006103516	0.649101257324219\\
nan	nan\\
3.65320014953613	1.189697265625\\
3.72599983215332	0.19329833984375\\
nan	nan\\
3.52409934997559	0.914199829101562\\
3.72599983215332	0.19329833984375\\
nan	nan\\
3.52409934997559	0.914199829101562\\
3.65320014953613	0.989299774169922\\
nan	nan\\
3.52409934997559	0.914199829101562\\
3.36400032043457	0.550399780273438\\
nan	nan\\
3.72599983215332	0.19329833984375\\
3.65320014953613	0.989299774169922\\
nan	nan\\
3.72599983215332	0.19329833984375\\
3.36020088195801	0.547897338867188\\
nan	nan\\
3.72599983215332	0.19329833984375\\
2.53399848937988	0.0407981872558594\\
nan	nan\\
3.72599983215332	0.19329833984375\\
2.6609001159668	0.084503173828125\\
nan	nan\\
3.72599983215332	0.19329833984375\\
3.36400032043457	0.550399780273438\\
nan	nan\\
0.624700546264648	0.644702911376953\\
0.186100006103516	0.649101257324219\\
nan	nan\\
3.36020088195801	0.547897338867188\\
2.6609001159668	0.084503173828125\\
nan	nan\\
3.36020088195801	0.547897338867188\\
3.36400032043457	0.550399780273438\\
nan	nan\\
2.53399848937988	0.0407981872558594\\
2.6609001159668	0.084503173828125\\
nan	nan\\
};

\addplot[only marks, mark=*, mark options={}, mark size=1.5811pt, color=mycolor1, fill=mycolor1] table[row sep=crcr]{%
x	y\\
2.13809967041016	3.70050048828125\\
2.64369964599609	3.32180404663086\\
1.00749969482422	3.67480087280273\\
1.63420104980469	3.56869888305664\\
2.64369964599609	2.69680023193359\\
2.13899993896484	2.69209671020508\\
1.63420104980469	2.78250122070312\\
0.961099624633789	2.69100189208984\\
1.63420104980469	2.24029922485352\\
3.19409942626953	2.09109878540039\\
2.64369964599609	1.85800170898438\\
2.13899993896484	1.99480056762695\\
1.1294994354248	2.46269607543945\\
1.81710052490234	1.68230056762695\\
1.63420104980469	1.95490264892578\\
3.14849853515625	1.77890014648438\\
2.64369964599609	1.33929824829102\\
2.13899993896484	1.50730133056641\\
1.1294994354248	1.90569686889648\\
1.63420104980469	1.11449813842773\\
3.14849853515625	1.59299850463867\\
3.2903995513916	1.68230056762695\\
2.16760063171387	0.676101684570312\\
0.35099983215332	1.59850311279297\\
1.1294994354248	1.29710006713867\\
1.63420104980469	0.993598937988281\\
3.14849853515625	1.12820434570312\\
3.65320014953613	1.09080123901367\\
3.6599006652832	0.994895935058594\\
2.56709861755371	0.394302368164062\\
1.1294994354248	0.64630126953125\\
1.63420104980469	0.106800079345703\\
3.65320014953613	1.189697265625\\
3.52409934997559	0.914199829101562\\
3.72599983215332	0.19329833984375\\
0.624700546264648	0.644702911376953\\
3.65320014953613	0.989299774169922\\
3.36020088195801	0.547897338867188\\
2.53399848937988	0.0407981872558594\\
2.6609001159668	0.084503173828125\\
0.186100006103516	0.649101257324219\\
3.36400032043457	0.550399780273438\\
};

\addplot[only marks, mark=*, mark options={}, mark size=1.5811pt, color=mycolor2, fill=mycolor2] table[row sep=crcr]{%
x	y\\
2.91889953613281	1.97455024719238\\
0.405400276184082	0.646902084350586\\
1.88660049438477	2.7372989654541\\
1.38185024261475	1.2057991027832\\
2.64369964599609	2.27740097045898\\
1.72565078735352	1.39839935302734\\
0.740249633789062	1.75209999084473\\
0.740249633789062	1.44780158996582\\
2.13899993896484	1.75105094909668\\
3.58864974975586	0.951749801635742\\
0.984299659729004	3.18290138244629\\
1.47329998016357	1.79399871826172\\
1.38185024261475	1.2057991027832\\
2.1385498046875	3.19629859924316\\
1.38185024261475	1.2057991027832\\
0.487850189208984	1.12160301208496\\
1.88660049438477	2.7372989654541\\
1.97805023193359	1.83855056762695\\
0.740249633789062	1.44780158996582\\
1.72565078735352	1.39839935302734\\
1.88615036010742	3.63459968566895\\
1.97805023193359	1.59480094909668\\
1.88660049438477	2.7372989654541\\
1.38185024261475	2.62259864807129\\
1.97805023193359	1.83855056762695\\
1.47329998016357	1.79399871826172\\
1.97805023193359	1.59480094909668\\
2.64369964599609	1.5986499786377\\
1.88660049438477	2.4661979675293\\
2.96364974975586	0.471099853515625\\
3.21944904327393	1.63764953613281\\
0.487850189208984	1.12160301208496\\
3.21944904327393	1.63764953613281\\
1.97805023193359	1.59480094909668\\
1.63420104980469	3.17560005187988\\
2.96364974975586	0.471099853515625\\
2.1385498046875	3.19629859924316\\
3.40084934234619	1.39134788513184\\
2.64369964599609	1.5986499786377\\
2.13899993896484	1.75105094909668\\
1.47329998016357	1.79399871826172\\
2.1385498046875	3.19629859924316\\
1.72565078735352	1.39839935302734\\
1.63420104980469	3.17560005187988\\
2.1385498046875	3.19629859924316\\
1.38185024261475	0.819950103759766\\
1.63420104980469	2.09760093688965\\
0.740249633789062	2.03059959411621\\
1.88660049438477	1.97485160827637\\
1.63420104980469	2.09760093688965\\
1.47329998016357	1.79399871826172\\
2.6053991317749	0.866800308227539\\
3.14849853515625	1.3606014251709\\
3.14849853515625	1.3606014251709\\
1.97805023193359	1.83855056762695\\
1.72565078735352	1.39839935302734\\
2.36734962463379	0.535202026367188\\
1.72565078735352	1.39839935302734\\
1.1294994354248	0.971700668334961\\
0.740249633789062	1.75209999084473\\
2.6053991317749	0.866800308227539\\
1.63420104980469	3.17560005187988\\
2.1385498046875	3.19629859924316\\
0.405400276184082	0.646902084350586\\
1.88660049438477	3.13039779663086\\
1.29765033721924	2.73675155639648\\
1.0452995300293	2.57684898376465\\
0.656049728393555	2.14475250244141\\
1.63420104980469	1.05404853820801\\
3.25434970855713	0.838050842285156\\
0.740249633789062	1.75209999084473\\
1.47329998016357	1.79399871826172\\
0.487850189208984	1.12160301208496\\
1.88660049438477	2.4661979675293\\
0.740249633789062	1.44780158996582\\
2.1385498046875	3.19629859924316\\
2.64369964599609	2.27740097045898\\
0.67924976348877	2.63665199279785\\
0.984299659729004	3.18290138244629\\
2.85779857635498	0.761253356933594\\
2.36734962463379	0.535202026367188\\
0.487850189208984	1.12160301208496\\
1.29765033721924	2.73675155639648\\
1.90090084075928	0.391450881958008\\
};

\end{axis}
\end{tikzpicture}%

%% file: framework.tex
We propose a framework that follows a local upsampling scheme. Therefore, the point cloud is partitioned into blocks. For the block partitioning the three-dimensional space is divided into dices of size $N\times N\times N$. Next, each point of the point cloud is assigned to one dice based on its location. However, the points remain on their real-valued original position. No voxelization is pursued. This holds the advantage that points that are near to each other in the three-dimensional space but are far away from each other in the saved structure of the point cloud data type are assigned to the same block and thus, can be computed together. The number of points in one block may vary. However, this might be a challenge for a learning-based approach but it is not a problem for our model-based approach. We then assume that all points in each block are element of the same surface which is a non-closed convex hull. This assumption allows us to describe the shape of the surface as a model. In the following, we are using a frequency-based model for the local representation of the object's surface as we could already upsample the color attribute successfully following a frequency-selective approach \cite{Heimann_2021}.\par
The assumption that the block's content is a valid function has to be confirmed first. Thus, the variances of the x-, y-, and z-component of the points' location are calculated, respectively. We assume that a plane can be represented in terms of a valid function if the model is built in the two dimensions showing the largest variances. The dimension showing the smallest variance is assumed to not show any closed form and thus, is valid to be represented by a function $q=f[o,p]$. 
If the surface should not fulfill our assumption that the dimension holding the smallest variance represents a non-closed function, it can be assumed that the error that is made during the model generation process is small. Hence, 
\begin{align}
q &:=  \underset{(x, y, z)}{ \mathrm{argmin}}\{\mathrm{Var[x]}, \mathrm{Var[y]}, \mathrm{Var[z]} \}, 
\end{align}
where \(\mathrm{Var}[\cdot]\) depicts the variance.
Now, a smooth surface can be fit into the block. The smooth surface can be interpreted as a function and thus, the function can be estimated using a model. We are going to approximate the function by a frequency model. This approach has been shown to be beneficial in a large range of different applications such as image reconstruction \cite{Kaup_2005, Schoeberl_2011}, quarter sampling approaches \cite{Seiler_2015}, and image resampling \cite{Koloda_2017}. Furthermore, it can also be used for upsampling the color attribute of point clouds \cite{Heimann_2021}.\par
\subsection{Frequency-Selective Geometry Upsampling}
\label{Sec:FSGU}
The Frequency-Selective Geometry Upsampling (FSGU) assumes the plane $f$ to be represented as a weighted superposition of two-dimensional basis functions $\varphi$, i.e.,
\begin{equation}
\label{Eq:image}
q=f[o,p] = \sum_{k, l \in \mathcal{K}} c_{k, l} \varphi_{k, l}[o,p],
\end{equation}   
where \(k\) and \(l\) denote the two-dimensional frequency indexes from the set of available basis functions \(\mathcal{K}\). As basis functions from the orthogonal discrete cosine transform (DCT) are used, the expansion coefficient $c_{k, l}$ can also be interpreted as transform coefficient. The aim of the model is to determine the impact of each frequency and thus, to determine the value of each expansion coefficient. Therefore, the model $g^{(\nu)}$ is built in an iterative manner and tries to fit the original signal as good as possible. In each iteration $\nu$, the selected basis function $\varphi$ with frequency indexes $u$ and $v$ with according estimated expansion coefficient is added to the model from the iteration before, i.e.,
\begin{equation}
\label{Eq:modelGeneration}
g^{(\nu)}[o,p] = g^{(\nu -1)}[o,p] + \hat{c}_{u, v} \varphi_{u, v}[o,p].
\end{equation}
The model $g^{(0)}$ is initialized to zero. The best fitting basis function in each iteration is found by minimizing the residual energy the most. Hence, the residual $r$ between the original signal and the model has to be determined in every iteration $\nu$ and is given as
\begin{equation}
\label{Eq:Residual}
r^{(\nu)}[o,p] = f[o,p] - g^{(\nu)} [o,p].
\end{equation}
Then, the residual energy 
\begin{equation}
E^{(\nu)} = \sum_{(o,p)} w[o,p]\left(r^{(\nu)}[o,p]\right)^2
\end{equation}
is calculated. The local weighting function $w[o,p]$ is defined as isotropic decaying window function and favors points close to the center. Finally, the best fitting frequency in iteration $\nu$ is selected by minimizing the residual energy in every iteration, i.e., 
\begin{equation}
{(u, v)} = \underset{{(k, l)}}{\mathrm{argmax}} \left( \Delta E_{k, l}^{(\nu)} w_{f}[k,l] \right).
\end{equation}
A spectral weighting function $ w_{f}[k,l]$ is incorporated in the selection process. Low frequency parts are favored as they are crucial for the representation of a smooth surface. Higher frequency parts tend to create a noisy surface which is not desired. Nevertheless, if a high frequency is dominant in the surface, it can be chosen and incorporated into our model. The iteration terminates if a predefined maximum number of iterations is met. Finally, a continuous model of the surface is established. This continuous function is then sampled for additional points in order to upsample the point cloud. The definition of the exact position of the additional sampling points in the \(o-p\)-plane is defined in the upcoming section. In Fig.~\ref{Fig:duckmodelinput} the continuous model and the original points are shown for one block of the \texttt{Duck} point cloud \cite{3DColMesh}. The original points are marked in blue. The model does not intersect with all points but approaches them sufficiently close. 

\subsection{Definition of Point Locations}
\label{sec:pointlocs}
For geometry upsampling, the location of the additional points has to be defined in the \(o-p\)-plane first. A straightforward way of defining the positions is to use a triangulation-based approach as in \cite{Alexa_2001, Dinesh_2019}. We pursue a Delaunay triangulation in the \(o-p\)-plane. The triangulation is shown in Fig.~\ref{Fig:inputoutput}. The blue points are the original points and thus, these are the corner points of the triangles. For our approach, we add points at the middle of each edge. Those are given in red. If the predefined scaling factor demands for a greater number of additional points, the triangulation is applied recursively. Thus, in a next step, the Delaunay triangulation would be pursued on all points, red and blue, from Fig.~\ref{Fig:inputoutput}. Once again, additional points would be inserted at the middle of each edge. \par
\begin{figure}[t!]
\vspace{-1.7cm}
\resizebox{\columnwidth}{!}{\input{figures/flow_chart}}
\caption{\label{Fig:flowchart} Flow chart of the proposed method.}
\end{figure}
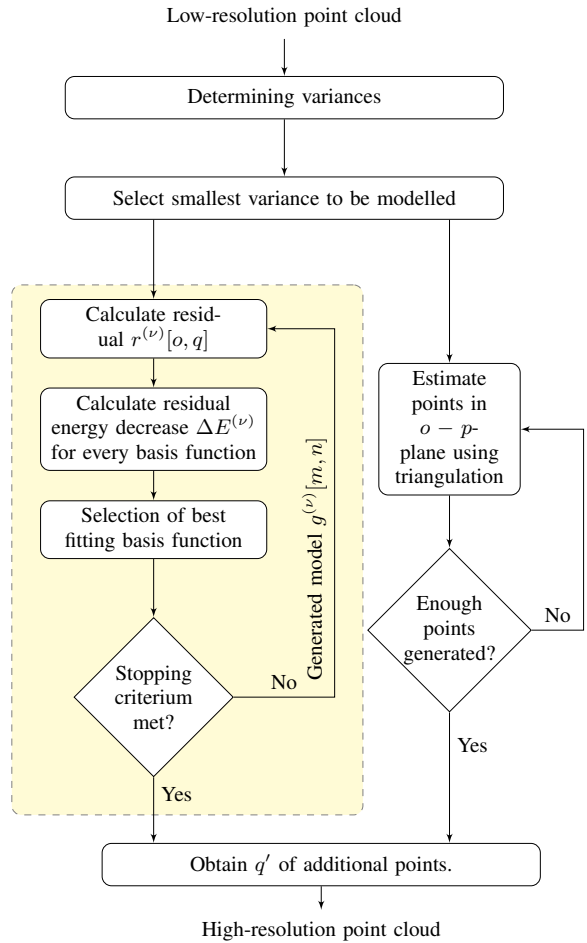

With the position of the inserted points \(o'\) and \(p'\) we can evaluate our estimated model 
\begin{equation}
\label{Eq:evaluationonmesh}
q'=f[o',p'] = \sum_{k, l \in \mathcal{K}} \hat{c}_{k, l} \varphi_{k, l}[o',q'].
\end{equation}   
In Fig.~\ref{Fig:duckmodelinputoutput}, the upsampled points are denoted in red. They fit well among the original points. A summary of the proposed framework is given as a flow chart in Fig.~\ref{Fig:flowchart}.

%% file: figures/flow_chart.tex

\definecolor{colorflow}{rgb}{0.00000,0.44700,0.74100}%
\makeatletter
\newif\iftikz@ortho@preflush
\tikz@ortho@preflushtrue
\let\tikz@origtotarget\pgfutil@empty
\tikzset{
  |-/.style={to path={|- (\tikztotarget) \tikztonodes}},
  -|/.style={to path={-| (\tikztotarget) \tikztonodes}},
  *|/.style={to path={%
    \pgfextra
      \iftikz@shapeborder
        \tikz@scan@one@point\pgfutil@firstofone(\tikztotarget)\relax
        \ifdim\pgf@y>\tikz@lasty\relax
          \edef\tikztostart{\tikztostart.north}%
        \else
          \edef\tikztostart{\tikztostart.south}%
        \fi
      \fi
    \endpgfextra
    (\tikztostart-|\tikztotarget) -- (\tikztotarget) \tikztonodes
  }},
  *-/.style={to path={%
    \pgfextra
      \iftikz@shapeborder
        \tikz@scan@one@point\pgfutil@firstofone(\tikztotarget)\relax
        \ifdim\pgf@x>\tikz@lastx\relax
          \edef\tikztostart{\tikztostart.east}%
        \else
          \edef\tikztostart{\tikztostart.west}%
        \fi
      \fi
    \endpgfextra
    (\tikztostart|-\tikztotarget) -- (\tikztotarget) \tikztonodes
  }},
  |*/.style={to path={%
    \pgfextra
      \tikz@scan@one@point\pgfutil@firstofone(\tikztotarget)\relax
      \iftikz@shapeborder
        \let\tikz@origtotarget\tikztotarget
        \ifdim\pgf@y>\tikz@lasty\relax
          \edef\tikztotarget{\tikztotarget.south}%
        \else
          \edef\tikztotarget{\tikztotarget.north}%
        \fi
      \fi
    \endpgfextra
    (\tikztostart) -- (\tikztostart|-\tikztotarget) \tikztonodes \ifx\tikz@origtotarget\pgfutil@empty\else\iftikz@ortho@preflush(\tikz@origtotarget)\fi\fi
  }},
  -*/.style={to path={%
    \pgfextra
      \tikz@scan@one@point\pgfutil@firstofone(\tikztotarget)\relax
      \iftikz@shapeborder
        \let\tikz@origtotarget\tikztotarget
        \ifdim\pgf@x>\tikz@lastx\relax
          \edef\tikztotarget{\tikztotarget.west}%
        \else
          \edef\tikztotarget{\tikztotarget.east}%
        \fi
      \fi
    \endpgfextra
    (\tikztostart) -- (\tikztostart-|\tikztotarget) \tikztonodes \ifx\tikz@origtotarget\pgfutil@empty\else\iftikz@ortho@preflush(\tikz@origtotarget)\fi\fi
  }},
  node as new start/.is if=tikz@ortho@preflush
  }

\tikzset{%
	back group/.style={fill=yellow!20,rounded corners, draw=black!50, dashed, inner xsep=12pt, inner ysep=11pt, yshift=-5pt}
}

\tikzstyle{decision} = [diamond, draw, fill=white!15, 
    text width=4.5em, text badly centered, node distance=3cm, inner sep=0pt]
\tikzstyle{block} = [rectangle, draw, fill=white!15, 
    text width = 20em, text centered, rounded corners, minimum height=2em] 
 \tikzstyle{block2} = [rectangle, draw=colorflow, fill=white!15, line width = 2pt,
    text width = 20em, text centered, rounded corners, minimum height=2em] 
 \tikzstyle{blockk} = [rectangle, draw, fill=colorflow, 
    text width = 20em, text centered, rounded corners, minimum height=2em] 
\tikzstyle{block3} = [rectangle, draw, fill=white!15, 
    text width = 10em, text centered, rounded corners, minimum width=10em, minimum height=2em] 
    \tikzstyle{block4} = [rectangle, draw, fill=white!15, 
    text width = 6em, text centered, rounded corners, minimum width=6em, minimum height=2em] 
 \tikzstyle{blockk} = [rectangle, draw, fill=colorflow, 
    text width = 20em, text centered, rounded corners, minimum height=2em] 
\tikzstyle{blockkk} = [rectangle, draw=white, fill=white!15, 
    text width = 20em, text centered, rounded corners, minimum height=2em] 
\tikzstyle{blocksummarize} = [rectangle, draw, fill=white!15, line style=dashed,
    text width = 20em, text centered, rounded corners, minimum height=2em] 
\tikzstyle{line} = [draw, -latex']
\tikzstyle{cloud} = [draw, ellipse,fill=white!20, node distance=5cm,
    minimum height=2em]
    
    \makeatother
\begin{tikzpicture}[node distance = 1.5cm, auto]
    \node [blockkk] (init) {Low-resolution point cloud};
    \node [block, below of=init, node distance = 1.2cm] (var) {Determining variances};
    \node [block, below of=var] (smallvar) {Select smallest variance to be modelled};
    
    \node [block3, below of=smallvar, left of= smallvar, node distance = 1.95cm] (residual) {Calculate residual $r^{(\nu)}[o,q]$};
    \node [block3, below of=residual] (energy) {Calculate residual energy decrease $\Delta E^{(\nu)}$ for every basis function};
    \node [block3, below of = energy] (selection) {Selection of best fitting basis function};
    \node [decision, below of=selection, node distance = 2.5cm] (stop) {Stopping criterium met?};
    \node [block4, below of = smallvar, right of=smallvar, right = .5em of residual] (2dims) {Estimate points in \(o-p\)-plane using triangulation};
    \node [block, below of= stop, right of = stop, node distance = 2.5cm] (mesh) {Obtain \(q'\) of additional points.};
    \node [decision, below of = 2dims] (enoughpoints) {Enough points generated?};
    \node [blockkk, below of=mesh, node distance = 1cm] (finish) {High-resolution point cloud};
    
    \coordinate[right of=residual] (a1);  
    \coordinate[right of=stop] (e1); 
    \coordinate[right of=enoughpoints] (e2);
    \coordinate[left of=stop](helper);

    \path [line] (init) -- (var);
    \path [line] (var) -- (smallvar);
    \path [line] (smallvar) to[*|]  (residual);
    \path [line] (residual)  -- (energy);
    \path [line] (energy) -- (selection);
    \path [line] (selection) -- (stop);
    \path [line] (stop) -| node [near start]{No}([xshift=1.2cm]e1) -- node[sloped, anchor=center, above, text width = 5cm](atest){Generated model $g^{(\nu)}[m,n]$}([xshift=1.2cm]a1) -- (residual);
    \path [line] (stop) to[|*] node[near start]{Yes} (mesh);
    \path [line] (mesh) -- (finish);
    \path [line] (smallvar) to[*|]  (2dims);
    \path [line] (2dims) -- (enoughpoints);
	\path [line] (enoughpoints) to[|*] node[near start](yesenough){Yes} (mesh);
     \path [line] (enoughpoints.east) -| node [near start](notenough){No}([xshift=.5cm]e2) |-  (2dims.east);
     
    \path (atest) +(0.3, 10) coordinate (a2);
    
    \begin{scope}[on background layer]
    		\node (bk2) [back group] [fit=(residual) (atest) (stop)] {};
    		\node[anchor=south east,inner sep=1pt,outer sep=1pt,opacity=0.5,font=\sffamily\bfseries,text=gray] at (bk2.south east){}; 
    		
    \end{scope}
    
    

\end{tikzpicture}

%% file: evaluation.tex
\begin{table*}[t!]
\caption{\label{Tab:results} Results  given in terms of Point to Point (P2Point) and Point to Plane (P2Plane) error~\cite{Tian_2017} and upsampling factors of 2 and 4, respectively. Avg gives the average results of the shown point clouds.}
\begin{tabular}{|l|l|l||c|c|c|c|c|c|c|c|c|c|c|c||c|} 
\hline				
\hspace{-.1cm}\parbox{.65cm}{\vspace{-.45cm}Scale\newline factor} 	& 	Metric 	&	Method & 	\rotatebox{85}{\texttt{Dragon}} & 	\rotatebox{85}{\texttt{Duck}} & 	\rotatebox{85}{\texttt{Jaguar}} 	& \rotatebox{85}{\texttt{Rabbit}} 	& 	\rotatebox{85}{\texttt{camel}} & 	\rotatebox{85}{\texttt{elephant}}  & \rotatebox{85}{\texttt{kitten}} 	& 	\rotatebox{85}{\texttt{star}} & \rotatebox{85}{\texttt{9}} & \rotatebox{85}{\texttt{21}} & \rotatebox{85}{\texttt{45}} & \rotatebox{85}{\texttt{76}} & Avg \\
\hline						
\multirow{6}{.4cm}{\(2\)} &  \multirow{3}{1.2cm}{P2Point $\times 10^{-3}$} 	& 	EAR \cite{Huang_2013} 	& 	11.3	& 	9.5	& 14.4		& 	29.6	&	13.6	& 	8.7	& 8.3		& 	7.4	& 	7.5 & 7.4		& 	7.9	& 	7.6	&11.1	\\ \cline{3-16}	
  				&					&	PU \cite{Yu_2018_PUNet}	& 	13.4	& 12.7		& 12.4	& 9.0 	&15.3 		&17.1 	&18.0	&22.4 		&13.3 	&13.6	& 14.0		&13.0 	& 14.5	\\ \cline{3-16}	
  				&					&	FSGU (Ours) & 	\textbf{1.9}	& \textbf{3.8}		& \textbf{1.5}		& \textbf{2.9}		&\textbf{3.5} 		&\textbf{3.6} 	& \textbf{3.6}	&\textbf{3.8}		&\textbf{3.1} 	& \textbf{2.5}	&\textbf{3.3} 		& \textbf{2.6}	& \textbf{3.0}	\\	
\cline{2-16}	
\cline{2-16}	
				& 			\multirow{3}{1.2cm}{P2Plane $\times 10^{-3}$} 	& 	EAR \cite{Huang_2013} 	& 	5.0	& 	4.3	& 	6.6	& 	9.2	&	2.8	& 	2.0	& 	1.7	& 	0.7	&3.0 	& 	3.6	& 	3.5	& 		4.0 & 3.9	\\ \cline{3-16}	
  				&					&	PU \cite{Yu_2018_PUNet}	& 	11.3 &	8.3 &	10.1 & 7.3&	7.5 &	8.6 &	7.0 & 7.6 &	7.1 &	 8.9 &9.4 &	8.6	& 8.5\\ \cline{3-16}	
  				&					&	FSGU (Ours) & 	\textbf{0.5} & \textbf{0.6} & \textbf{0.4} & \textbf{1.3}	& \textbf{0.5} & \textbf{0.5} & \textbf{0.4}	& \textbf{0.5} & \textbf{0.5} & \textbf{0.5}& \textbf{0.6} & \textbf{0.4}	 & \textbf{0.6} \\	
\hline	
\hline
\multirow{6}{.4cm}{\(4\)} &  \multirow{3}{1.2cm}{P2Point $\times 10^{-3}$} 	& 	EAR \cite{Huang_2013} 	& 27.8 & 12.7	& 16.6	& 12.1 &	17.9& 	20.6& 	16.4& 	15.2&  18.9& 14.6 	& 14.0	&	14.3 & 16.8\\ \cline{3-16}	
  				&					&	PU \cite{Yu_2018_PUNet}	& 	11.7 & 11.4 & 11.4 & 8.7 & 16.6 & 16.6 & 16.7 & 16.6 & 12.9 & 12.9 & 14.0 & 12.6	&  13.5	\\ \cline{3-16}	
  				&					&	FSGU (Ours) & \textbf{1.5} & \textbf{3.5} & \textbf{1.2} & \textbf{2.8} & \textbf{3.4} & \textbf{3.5} & \textbf{3.8} & \textbf{4.0} & \textbf{3.7} & \textbf{3.0} & \textbf{4.1} & \textbf{3.0}	& \textbf{3.1}\\	
\cline{2-16}	
\cline{2-16}	
				& 			\multirow{3}{1.2cm}{P2Plane $\times 10^{-3}$} 	& 	EAR \cite{Huang_2013} 	& 9.0	& 5.0	& 6.6	& 3.7 & 4.0 & 5.4	& 4.5	& 4.5	& 6.3 & 5.4 & 4.9	& 5.8& 5.4	\\ \cline{3-16}	
  				&					&	PU \cite{Yu_2018_PUNet}	&  6.0 & 6.3 & 6.5 & 4.5 & 6.4 & 6.8 & 7.2 & 7.3 & 5.8 & 6.1 & 6.4 & 7.2 & 6.4\\ \cline{3-16}	
  				&					&	FSGU (Ours) & 	\textbf{0.4} & \textbf{0.4} & \textbf{0.4} & \textbf{1.4} & \textbf{0.4} & \textbf{0.4} & \textbf{0.4} & \textbf{0.5} & \textbf{0.5} & \textbf{0.6} & \textbf{0.5} & \textbf{0.5}	& \textbf{0.5}\\	
\hline  
\end{tabular}
\end{table*}
We conduct extensive experiments to analyze the performance of our proposed FSGU framework. For the performance evaluation, we compare results of our proposed FSGU framework with two commonly used algorithms following two different approaches, the normal vector-based Edge-Aware Resampling (EAR) \cite{Huang_2013} and the data-driven learning-based neural network PU-Net~\cite{Yu_2018_PUNet}. Furthermore, we show results for the point clouds \texttt{Dragon, Duck, Jaguar} and \texttt{Rabbit} from \cite{3DColMesh}, \texttt{camel, elephant, kitten} and \texttt{star} from \cite{Yu_2018_PUNet} and \texttt{9, 21, 45} and \texttt{76} from \cite{Lian_2011, Pickup_2014}. We evaluate in terms of point-to-point (P2Point) and point-to-plane (P2Plane) error following the evaluation scheme of Tian et al. \cite{Tian_2017}. The P2Point error is determined as the normalized sum of the error vectors being the smallest distance between the points in the reference and the upsampled point cloud. For the P2Plane error the error vectors are projected along the direction of the normals and are summed up and normalized~\cite{Tian_2017}. \par
Furthermore, we normalize the point clouds to a cube with a length of side of one. For PU-Net, we use the pretrained net from the authors \cite{Yu_2018_PUNet}. Also the EAR implementation is the algorithm provided by the authors \cite{Huang_2013}. Table~\ref{Tab:results} shows the errors of the examined point clouds for two scaling factors. The last column gives the average error of the twelve incorporated point clouds in the evaluation. For the results in the upper part a scaling factor of \(2\) is incorporated, i.e., the number of points of each point cloud was doubled. For the lower part, the number of points in each point cloud was quadrupled. This corresponds to a scaling factor of \(4\) for which the PU-Net~\cite{Yu_2018_PUNet} was originally trained. The best results are given in bold. The relative behavior of the examined methods is similar for both scaling factors. FSGU performs best for all evaluated point clouds for both P2Point and P2Plane errors. This is independent of the used dataset as it can be observed for all point clouds. On average, FSGU reaches a P2Point error of \(3.0\times10^{-3}\) for the incorporated point clouds and scale factor \(2\). The second best performing technique is EAR with an average P2Point error of \(11.1\times10^{-3}\). Thus, the P2Point error of EAR is by a factor \(3.7\) worse than the P2Point error of FSGU. PU-Net performs worst in this evaluation. In terms of P2Plane error, the general behavior of the three methods is analog. PU-Net performs worst, followed by EAR and the best performing FSGU. The P2Plane error of EAR is by a factor of \(6.5\) worse than the error of FSGU.\\ 
For a scale factor of \(4\), FSGU is the best performing technique again. The average P2Point error slightly increases to \(3.1\times10^{-3}\), whereas the P2Plane error shows a slight decrease to \(0.5\times10^{-3}\). In terms of P2Point error, PU-Net is now the second best performing method. Here, FSGU shows a \(4.4\) times smaller error than PU-Net. EAR inserts the additional points outside the original hull of the object and thus, the P2Point error increases. \par   
\begin{figure}
\vspace{-.15cm}
\begin{subfigure}[t]{\columnwidth}
\includegraphics[scale=.2, trim=190 135 220 130 mm, clip=true]{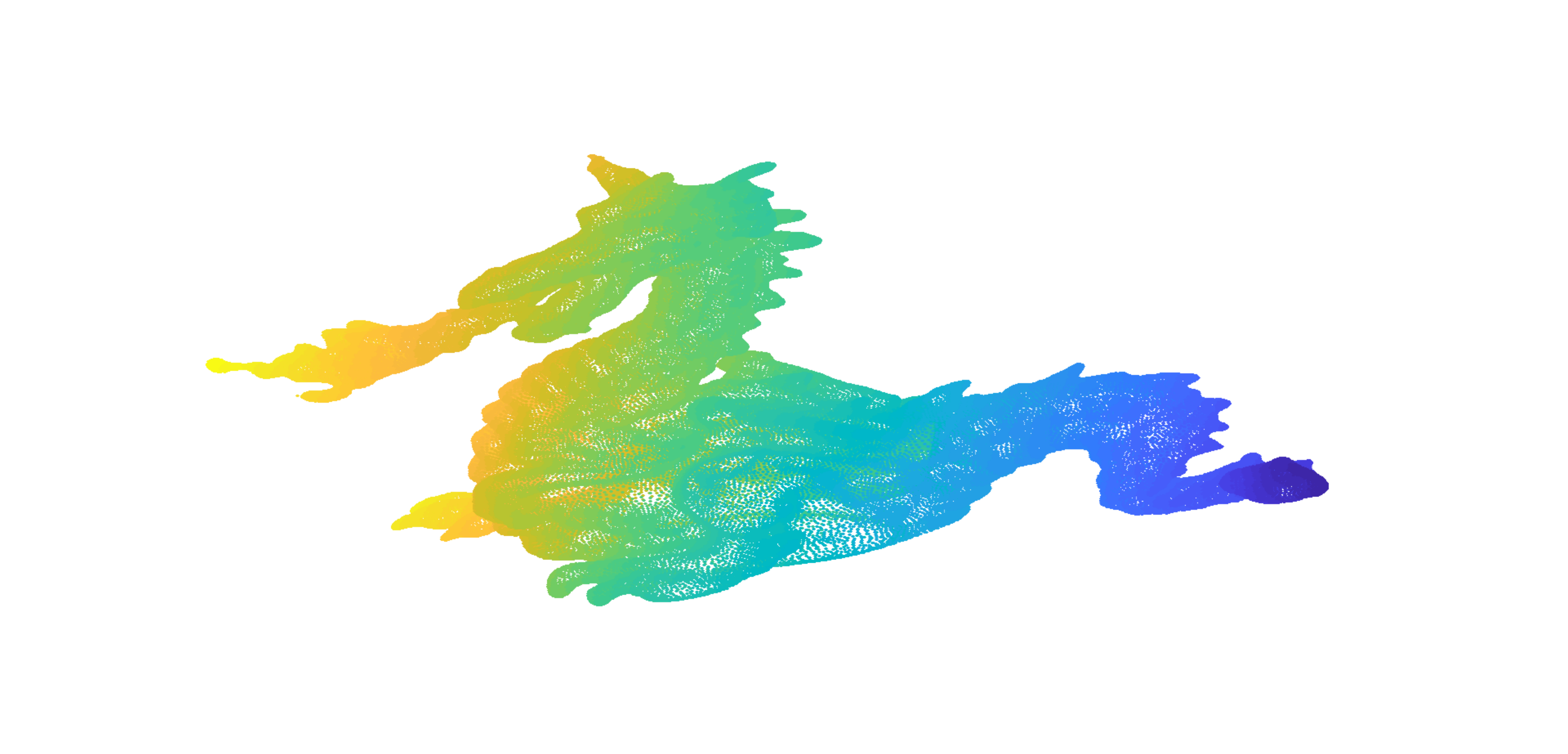}
\caption{\label{fig:dragonorig}Original.}
\end{subfigure}
\qquad
\begin{subfigure}[t]{\columnwidth}
\includegraphics[scale=.2, trim=190 135 220 130 mm, clip=true]{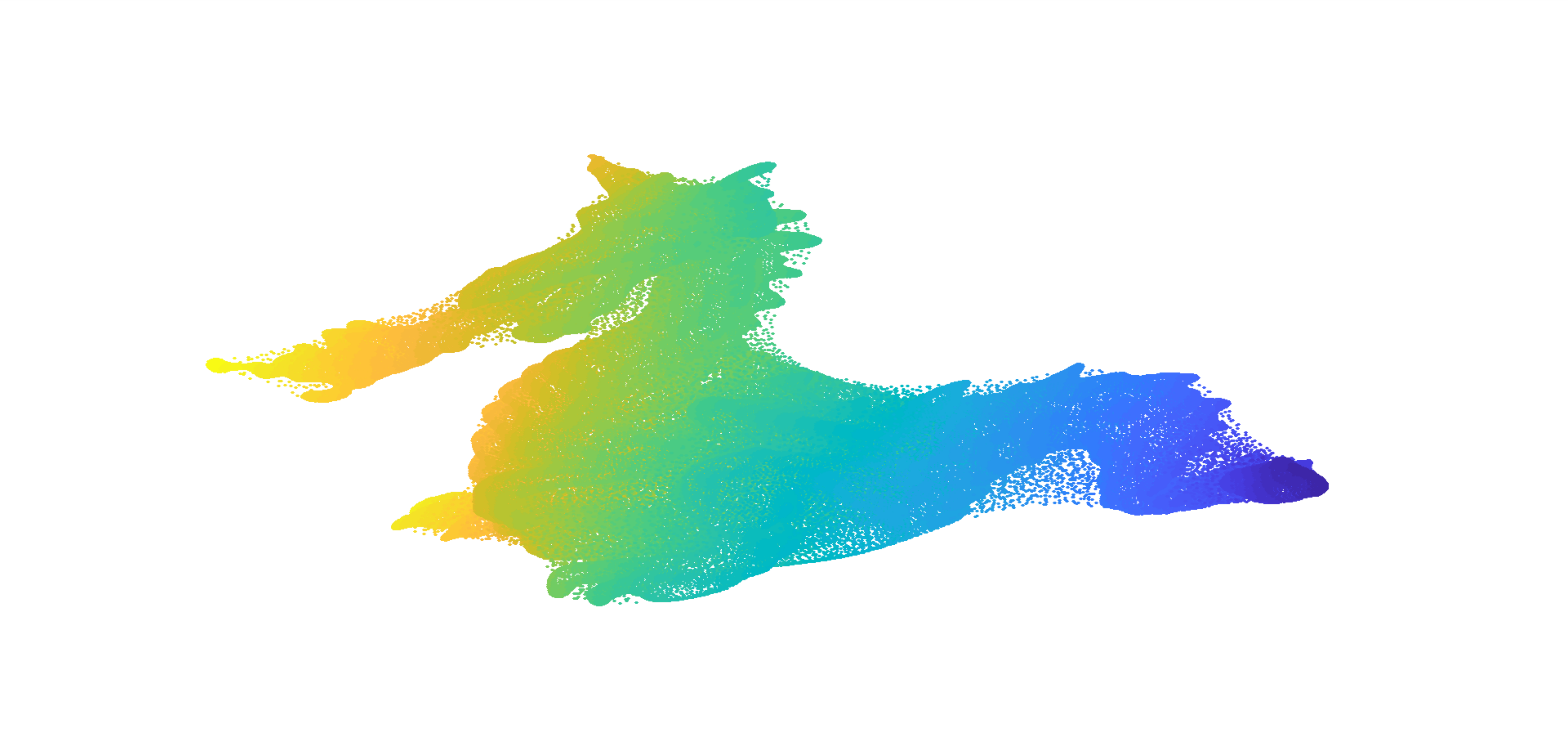}
\caption{\label{fig:dragonear}EAR \cite{Huang_2013}.}
\end{subfigure}
\qquad
\begin{subfigure}[t]{\columnwidth}
\includegraphics[scale=.2, trim=190 135 220 130 mm, clip=true]{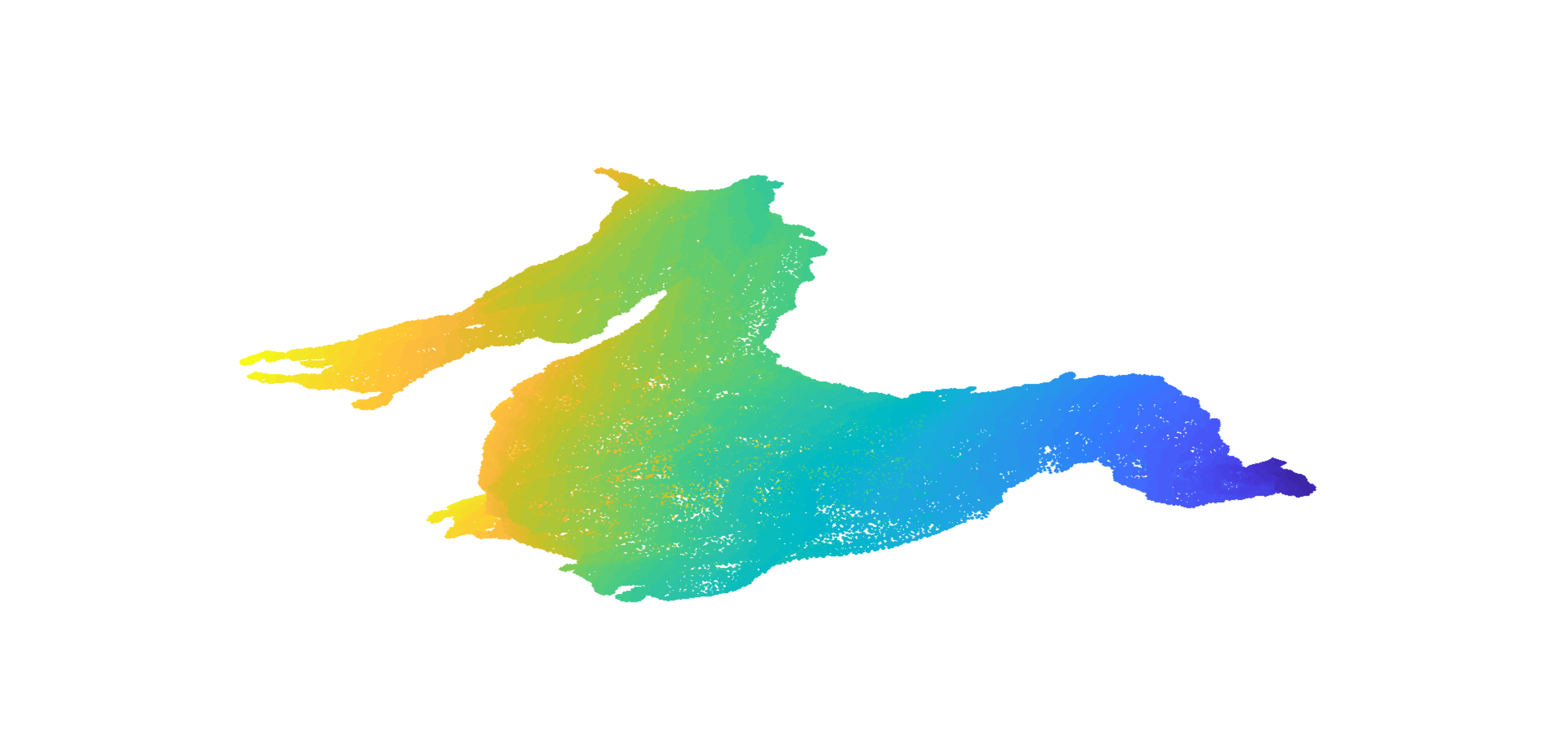}
\caption{\label{fig:dragonpu}PU-Net \cite{Yu_2018_PUNet}.}
\end{subfigure}
\qquad
\begin{subfigure}[t]{\columnwidth}
\includegraphics[scale=.2, trim=190 135 220 130 mm, clip=true]{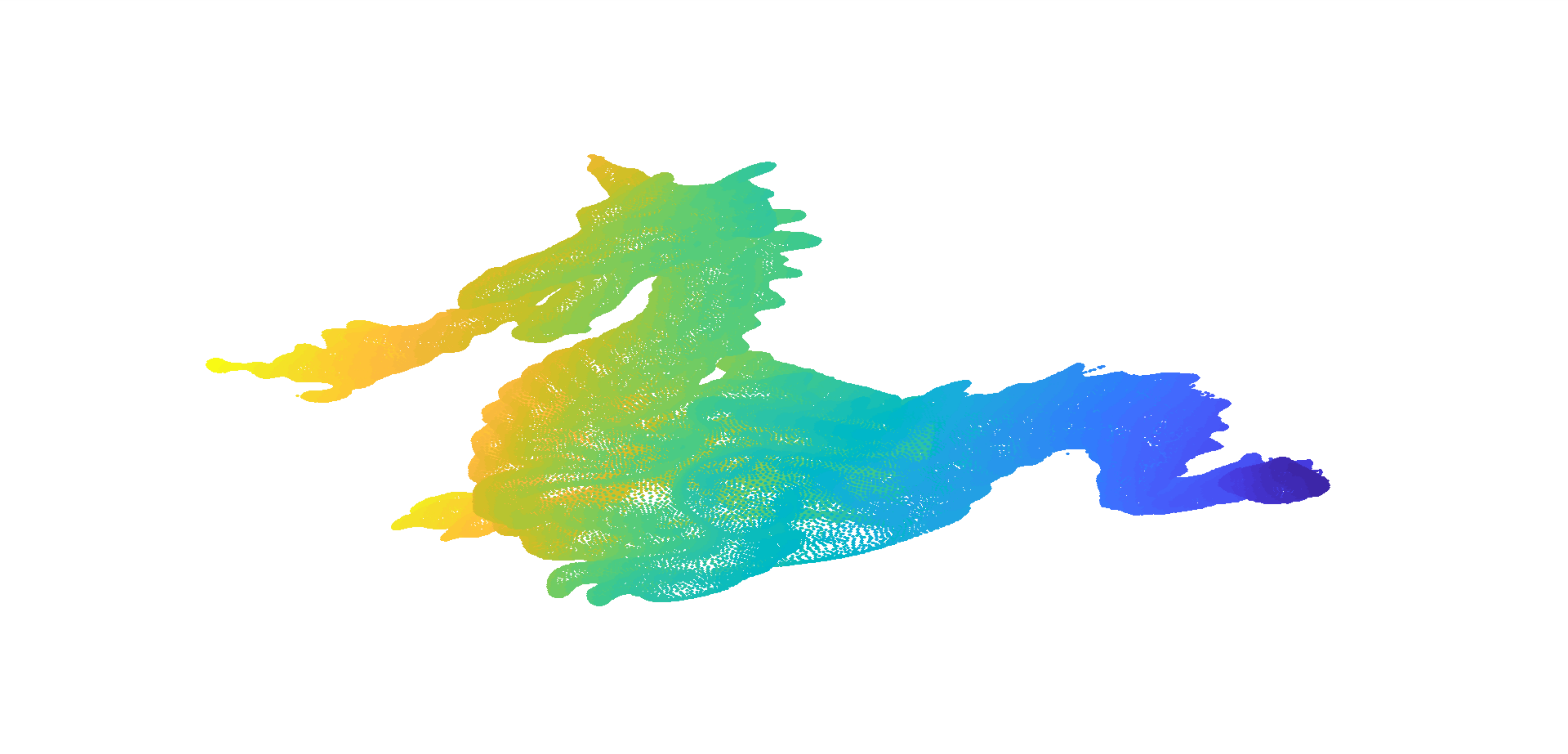}
\caption{\label{fig:dragonfsgu}FSGU (Ours).}
\end{subfigure}
\caption{\label{Fig:visexample} \texttt{Dragon} point cloud. Upsampling factor is 2. The depth is color coded for better interpretation.}
\vspace{-.3cm}
\end{figure}
In order to further demonstrate the differences of the three approaches, a visual example for the \texttt{Dragon} point cloud is given in Fig.~\ref{Fig:visexample}. The varying color of the point clouds emphasizes the depth component of the point clouds. The original input point cloud is given in Fig.~\ref{fig:dragonorig}. In Fig.~\ref{fig:dragonear}, the upsampled point cloud using EAR is shown. The upsampled result using PU-Net is given in Fig.~\ref{fig:dragonpu}. The final result incorporating our proposed FSGU scheme is shown in Fig.~\ref{fig:dragonfsgu}. The original point cloud appears more transparent compared to the upsampled point clouds as it incorporates only half of the points of the upsampled point clouds. The EAR method fails to model details of the dragon such as the bend of the tail. It rather models the rough form of the tail and fills empty regions in between the bends with additional points. It is also difficult for PU-Net to add additional points to the tail of the dragon. Neither the peaks on top of the tail nor the filigree rhombus formed tip of the tail are upsampled satisfactorily. Our proposed FSGU models the tail accurately. The bend of the tail shows high accuracy and the rhombus at the tip of the tail is depicted clearly. Just some minor artifacts are visible in the middle of the bending tail. \par